\documentclass[manuscript]{aastex}

\def\degpnt{^{\circ}\kern-1.7mm.\kern+.35mm}
\def\arcpnt{"\kern-1.7mm.\kern+.35mm}
\def\minpnt{'\kern-1.0mm.\kern+.30mm}

\slugcomment{Submitted to AJ}

\shorttitle{Novae in Local Group Dwarfs}
\shortauthors{Neill \& Shara}

\begin{document}


\title{A Possible High Nova Rate for Two Local Group Dwarf Galaxies: \\ 
	M32 and NGC 205}
	

\author{James D. Neill}
\affil{Department of Physics and Astronomy, University of Victoria,
	Elliott Building, 3800 Finnerty Road, Victoria, BC, V8P 5C2, Canada}
\email{neill@uvic.ca}

\author{Michael M. Shara}
\affil{American Museum of Natural History, 79th and Central Park West
	New York, NY, 10024}
\email{mshara@amnh.org}



\begin{abstract}

We report the results of a preliminary nova survey of Local Group dwarf
ellipticals.  We used the 15' field-of-view CCD camera on the 0.8 m
telescope at the Tenagra Observatory to observe M32, NGC 205, NGC 147,
and NGC 185 in their entirety every clear night over a 4.5 month interval
and discovered one nova in M32 and a candidate nova in NGC 205.  The nova in 
M32 was
verified spectroscopically.  The nova candidate in NGC 205 had an
unusually low peak luminosity (M$_V = -5.1$), and we were unable to obtain
spectroscopic verification.  Archival {\it HST} images provide us with a
limit on the outburst amplitude for this object of $>$4.6 Vmag.  These 
facts prompt us to consider the possibility
that this object is not a genuine nova.  We report a high bulk nova rate
for M32 of $2^{+2.4}_{-1.0}$ yr$^{-1}$ and, assuming the candidate nova is
correctly identified, for NGC 205 of $2^{+2.2}_{-1.0}$
yr$^{-1}$.  If the NGC 205 variable is not a nova, we calculate an upper
limit on the bulk nova rate for NGC 205 of 1.5 yr$^{-1}$.  We report upper 
limits on the bulk nova rates in NGC 147 of 2 yr$^{-1}$ and
NGC 185 of 1.8 yr$^{-1}$ and a combined bulk nova rate
for the four galaxies of $4^{+4.2}_{-1.4}$ yr$^{-1}$ (2$^{+3.9}_{-1.4}$
yr$^{-1}$ without the NGC 205 nova candidate).  The bulk rates
we report here are based on Monte Carlo simulations using nova maximum
magnitudes and decline rates and individual epoch frame limits.  From the
Monte Carlo rates, integrated and extinction corrected V-band photometry,
and (V-K)$_0$ colors we derive a luminosity specific nova rate for
M32 of 12.0$^{+14.4}_{-6.0}$ yr$^{-1} [10^{10} L_{\odot,K}]^{-1}$ and for
NGC 205 of 29.3$^{+32.3}_{-14.7}$ yr$^{-1} [10^{10} L_{\odot,K}]^{-1}$
and for the combined 4 galaxies of 14.1$^{+14.8}_{-4.9}$ yr$^{-1}
[10^{10} L_{\odot,K}]^{-1}$ (7.0$^{+13.7}_{-4.9}$ yr$^{-1}[10^{10}
L_{\odot,K}]^{-1}$ without the NGC 205 nova candidate).  The higher combined 
rate is 2.5$\sigma$ higher than expected from assuming a constant 
luminosity specific nova rate as a function of K-band luminosity as derived
from more massive galaxies.  If the higher rate is confirmed
by surveys in subsequent seasons, it would imply that either 
dwarf ellipticals have a higher interacting binary fraction than their 
higher mass counter parts, or that the completeness is higher for these
less complex systems and the nova rates for larger, more distant 
systems are systematically underestimated.

\end{abstract}




\section{Introduction}

Extragalactic novae are potentially important as tracers of
close binary stars in other galaxies.  
Current estimates of the bulk nova rates in galaxies such as M31 
\citep{sha01} of $\sim$37 yr$^{-1}$ imply that in a few years the number of
observed novae could be quite large for such a system.  Novae
represent an important complement to the increasing data on extragalactic
X-ray binaries afforded by {\it Chandra} and {\it XMM}.  The brightness of 
the nova outburst
(M$_V$ of $-6$ to $-10$ at maximum) betrays their presence to beyond
the Virgo cluster with current telescopes.  By observing extragalactic
novae it is possible to trace the frequency and distribution of the
close binaries that produce them in many extragalactic environments,
thus allowing an exploration of close binary populations and the factors
that influence their formation.

One of the most basic investigations into these factors is to plot the
normalized nova rate versus the luminosity of the host galaxy and see
if any trend can be detected.  Various versions of this plot have been
produced over the years \citep{del94,sha00,fer03}, but systematic effects
continue to dominate the published nova rates.  We have found that nova
rates are subject to biases that tend to underestimate the bulk rate for
a given galaxy \citep{nei04}.  Most severe of these biases is the one
imposed by telescope scheduling, which, until recently, provided only
short, widely spaced runs for sampling nova rates in external galaxies.

We have attempted to overcome this bias by using a dedicated telescope
to observe the target galaxy in its entirety, every clear night for many
months.  Our first survey of this type was of M81 \citep{nei04} and
produced a bulk nova rate 40\% higher than previous studies \citep{sha00}.
However, we also demonstrated the effects of dust in the disk on the
detection of novae in the bulge, implying that our higher bulk rate could be
still be low by up to a factor of 2.

Nearby dwarf ellipticals offer relatively dust-free targets that could
potentially be surveyed for novae with close to 100\% completeness.
Yet, if one examines the plot of normalized nova rate versus galaxy luminosity,
such as presented in the references above, it is clear that at the low
luminosity end, there is much uncertainty.  This is for obvious reasons.
In particular, low luminosity systems produce few novae per year and
so the sample is small.  The low luminosity systems must be nearby and
so are often very large and difficult to survey in their entirety (e.g.
M33, LMC, SMC).

We took advantage of the availability of telescope time on an hourly
basis, provided by the Tenagra observatory, to perform a comprehensive,
nightly survey of four local group dwarf galaxies with the aim of
refining the nova rates at the low luminosity end.  We surveyed M32,
NGC 205, NGC 147, and NGC 185 for over four months every clear night.
We are also surveying the LMC with a different telescope, and will 
present the results from that survey in a subsequent paper.

These surveys will continue for several years and will provide accurate
nova rates for the low luminosity systems, allowing us to determine
if there is indeed a trend in nova rate with luminosity.  In order to
constrain binary formation and evolution theory, this kind of survey must
be accompanied by comprehensive, densely time-sampled surveys of higher
mass galaxies.  Only by removing systematic biases can we determine if
there is a universal nova rate per unit luminosity, or if the nova rate
is influenced by the mass of the host galaxy.

\section{Observations}

We used the SITe based 1024x1024 pixel CCD camera on the Tenagra
0.8 m telescope for our observations of the local group dwarfs.
This configuration yields a pixel scale of 0\arcpnt87 px$^{-1}$ and a
field size of 15' on a side, allowing us to cover each galaxy in its
entirety for each epoch of observation.

The majority of the survey observations were taken through a standard
Johnson V filter.  This filter was chosen to maximize the sensitivity of
the telescope and detector combination.  Once a nova was discovered,
we initiated additional observations through the standard Johnson
B filter to derive nova colors.  Each individual exposure was 300s,
except for M32 which required a shorter exposure time of 150s to avoid
saturating the nuclear region.  We attempted to have 15 minutes total
exposure time per epoch.  Most epochs reached this goal, with only a
few having less exposure time.  The seeing for our observations had a
median of 2\arcpnt5 and ranged from 1\arcpnt7 to 4\arcpnt5.

Our survey ran from October 04, 2003 (JD 2452916.8) to February 18, 2004
(JD 2453053.6) covering a total of 136.8 days.  An additional epoch in
the I-band was generously obtained for us by John Thorstensen using the
Echelle direct CCD camera on the Hiltner 2.4m telescope.  In addition,
he was able to obtain two spectra using the Modspec on the 2.4m on JD
2453017.60 and two spectra on JD 2453022.60, allowing us to confirm the
nature of the M32 nova.  We were also fortunate that {\it HST} images
were available in the archive of both nova positions\footnote{
Observations made with the NASA/ESA Hubble Space Telescope, obtained from
the data archive at the Space Telescope Science Institute.  STScI is
operated by the Association of Universities for Research in Astronomy,
Inc. under NASA contract NAS 5-26555.}.  For M32 nova 1 we used WFPC2
images taken on JD 2449622 (proposal ID 5464, PI Rich).  For NGC 205
nova candidate 1 we used an ACS WFC image taken on JD 2452525 (proposal ID 9448,
PI Ferrarese).  We also used a 4m MOSAIC image from the NOAO science
archive taken on JD 2452528 (PI Massey).  Table~\ref{tab_obs} summarizes
all the observations presented in this paper.

\section{Reductions}

All Tenagra exposures were bias subtracted and flat fielded using the
standard tools in IRAF \citep{tod86}.  The exposures for a given epoch
(6 for M32, 3 for the others) were then registered and combined to
produce a coadded image for each epoch using the following DAOPHOT
programs \citep{ste87}: DAOPHOT to measure the point sources, DAOMATCH
and DAOMASTER to derive and refine the transformations, and MONTAGE2
to perform the registration and coaddition.  The registration master
was chosen to be the best coadded image from the entire set for a given
galaxy based on measurements of the seeing in each coadded image over the
entire survey.  The coaddition process removed all but a few cosmic rays.

The HST images were downloaded from the archive in reduced form.
The only processing that was required was to coadd the WFPC2 images to remove
cosmic rays.  The images were already registered, so this task was easily
achieved using the STSDAS {\it wfpc2} combine task.  The image from the NOAO
science archive required no processing.

The spectra were extracted using the IRAF apsum task, wavelength
calibrated using calibration lamps and night sky lines, and flux
calibrated using flux standard observations with the same configuration.
The wavelengths are accurate to better than 1\AA, but the fluxes are
probably only good to about 20\% due to unknown slit losses.

\section{Nova Detection \label{novdet}}

The coadded Tenagra images were blinked against each other to detect
changing point sources.  Nova candidates were required to be observed in
at least two epochs {\it and} to be missing on an epoch of sufficient
depth coverage to confirm its transient nature.  We also used the raw
images for an epoch to confirm the presence of the candidate in each
individual frame.  As a further verification, we checked the images from
the archives listed in Table~\ref{tab_obs}, and images from the Digitized
Sky Survey\footnote{ The Digitized Sky Survey was produced at the Space
Telescope Science Institute under U.S. Government grant NAG W-2166. The
images of these surveys are based on photographic data obtained using the
Oschin Schmidt Telescope on Palomar Mountain and the UK Schmidt Telescope.
The plates were processed into the present compressed digital form with
the permission of these institutions.}.

For the regions of our target galaxies near the nuclei, where the
intensity gradient makes detection more difficult, we used the spatial
filtering technique described in \citet{nei04}, allowing us to detect
novae to within 10" of the nuclei of M32 and NGC 205 and to within 2"
of the centers of NGC 147 and NGC 185.  This subtraction technique was
performed on each coadded image after which they were blinked against
each other.  For each coadded image of each galaxy we determined the
frame limit by using artificial stars and the exact techniques outlined
above for detecting the novae.

\begin{figure}
\plotone{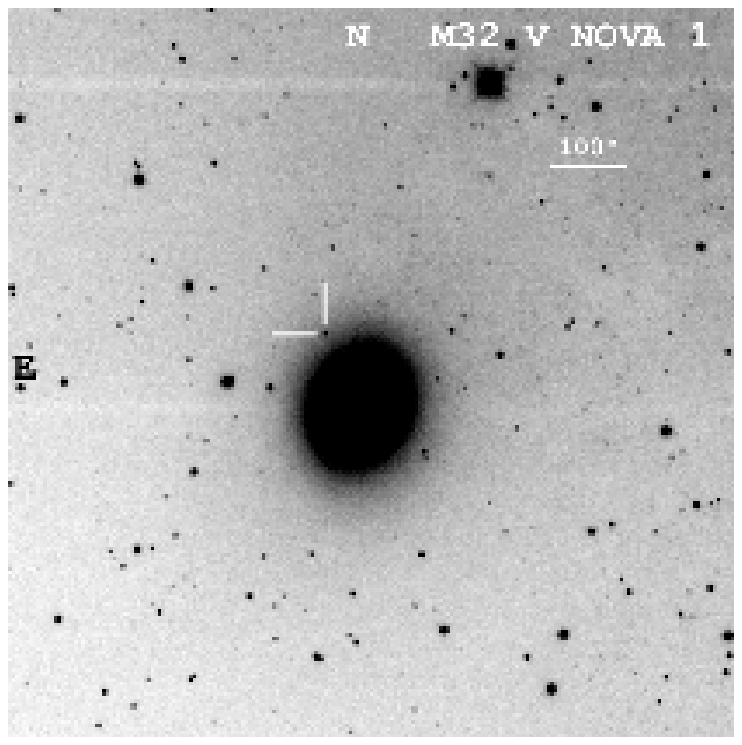}
\caption{Tenagra V-band image of M32 nova 1 on JD 2453009.62.  North is up and 
East to the left.  The nova is indicated by the two lines at right angles.
A scale bar of 100 arcseconds is also shown.
\label{m32n1}
}
\end{figure}

\begin{figure}
\plotone{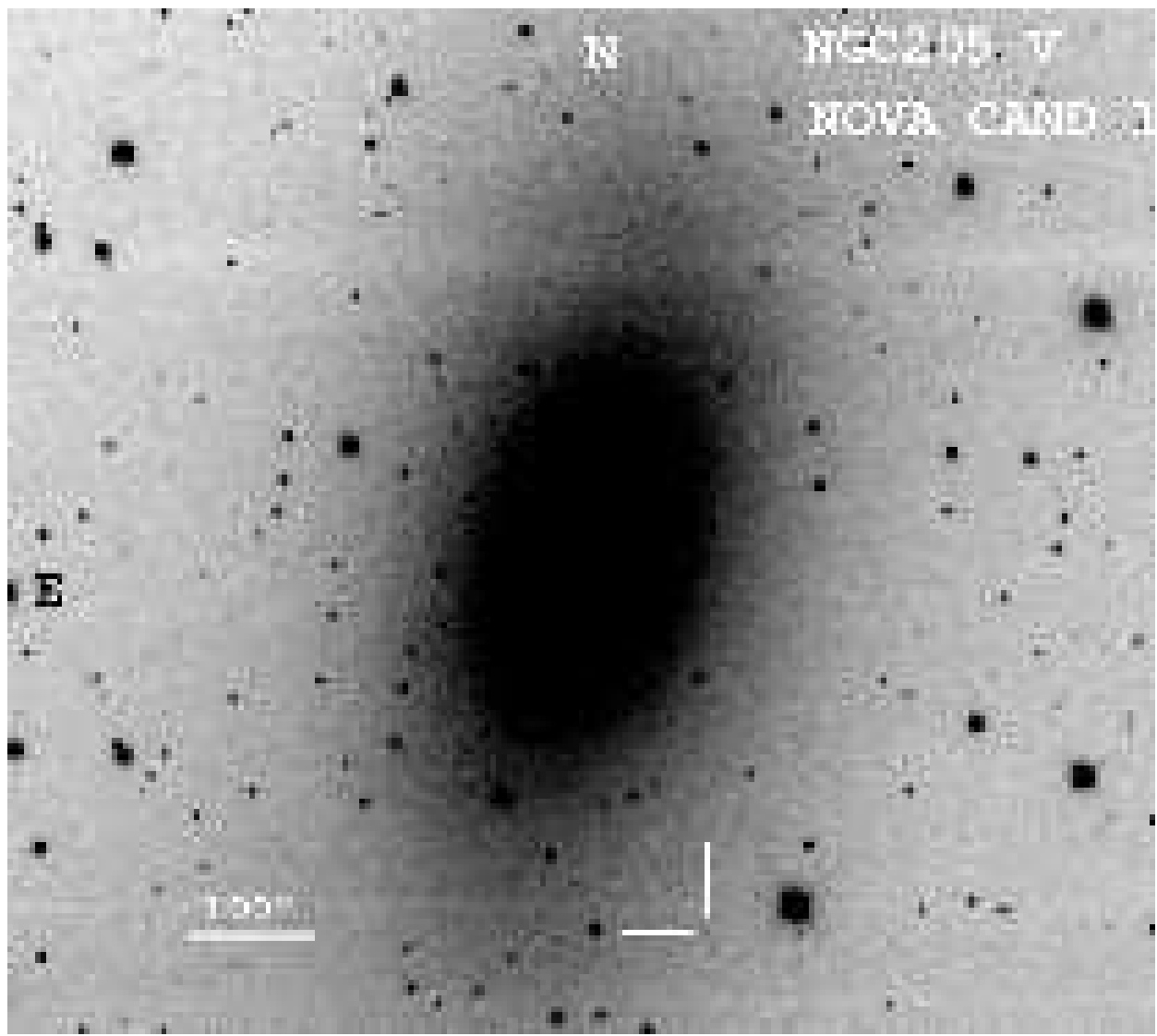}
\caption{Tenagra V-band image of NGC 205 nova candidate 1 on JD 2453017.64.  North is up 
and East to the left.  The nova candidate is indicated by the two lines at right 
angles.  A scale bar of 100 arcseconds is also shown.
\label{n205n1}
}
\end{figure}

Table~\ref{tab_nov_pos} gives the positions and number of detections
for the novae discovered in this survey.  The nova in M32 was discovered
first and is shown in outburst in Figure~\ref{m32n1}.  The nova candidate
in NGC 205 is shown in outburst in Figure~\ref{n205n1}.

\section{Nova Photometry}

Since crowding was not an issue, DAOPHOT aperture photometry was used
to measure point source brightnesses in each coadded image.  Variable
seeing was accounted for by setting the measurement aperture radius
in each image to 1/2 the FWHM of the stellar profile to maximize the
signal-to-noise ratio.  The FWHM was measured from a set of well exposed,
isolated stars in each image.

Calibration required the use of a diverse set of references.  For M32,
we used the study of \citet{mag92} which presents BVRI CCD photometry for
361,281 objects in the field of M31 and also includes M32.  For NGC 205,
the BVRI photometry published in \citet{lee96} was used.  NGC 147 and
NGC 185 were both calibrated using V-band photometry presented in the
study of \citet{now03}.  These studies all made corrections for galactic
extinction.  In all cases, the epoch with the best photometric conditions
was calibrated using stars in common with the references above and then
all other epochs were calibrated to the reference epoch.  The number
of objects used in the calibrations were as follows: over 100 for M32,
8 for NGC 205, 21 for NGC 147, and 19 for NGC 185.  In all cases we achieved
a photometric accuracy of 0.1 magnitude or better in all filters.

We measured stars in the {\it HST} images using very small apertures 
(1.6px for the WFPC and 1.08px for the ACS).  This avoided contamination
from cosmic rays (in the ACS image) and contamination due to the crowding 
at faint magnitudes for both of these images.  Instead of determining 
aperture corrections and using the standard photometric 
calibration, we chose to use calibrated reference stars to bootstrap the
calibration of our 
observed magnitudes.  We used the same references we used to calibrate the
Tenagra images, with considerably fewer objects, due to their smaller fields
of view.  For M32, only one of
the reference stars was available for photometric calibration, while for
NGC 205 six stars were available.  The NGC 205 calibration has an RMS
scatter of 0.2 magnitudes in the V-band.  We assume that the calibration
for M32 has a similar error.  We checked our calibration against the
photometric calibrations from the {\it HST} image headers and they agree to
within the uncertainties in determining the aperture corrections.

Table~\ref{m32n1_phot} and Table~\ref{n205n1_phot} present our calibrated
photometry for both objects at each observed epoch.  The errors presented
are 1$\sigma$ internal photometric errors.

\section{The Light Curves}

\begin{figure}
\includegraphics[angle=90,scale=.70]{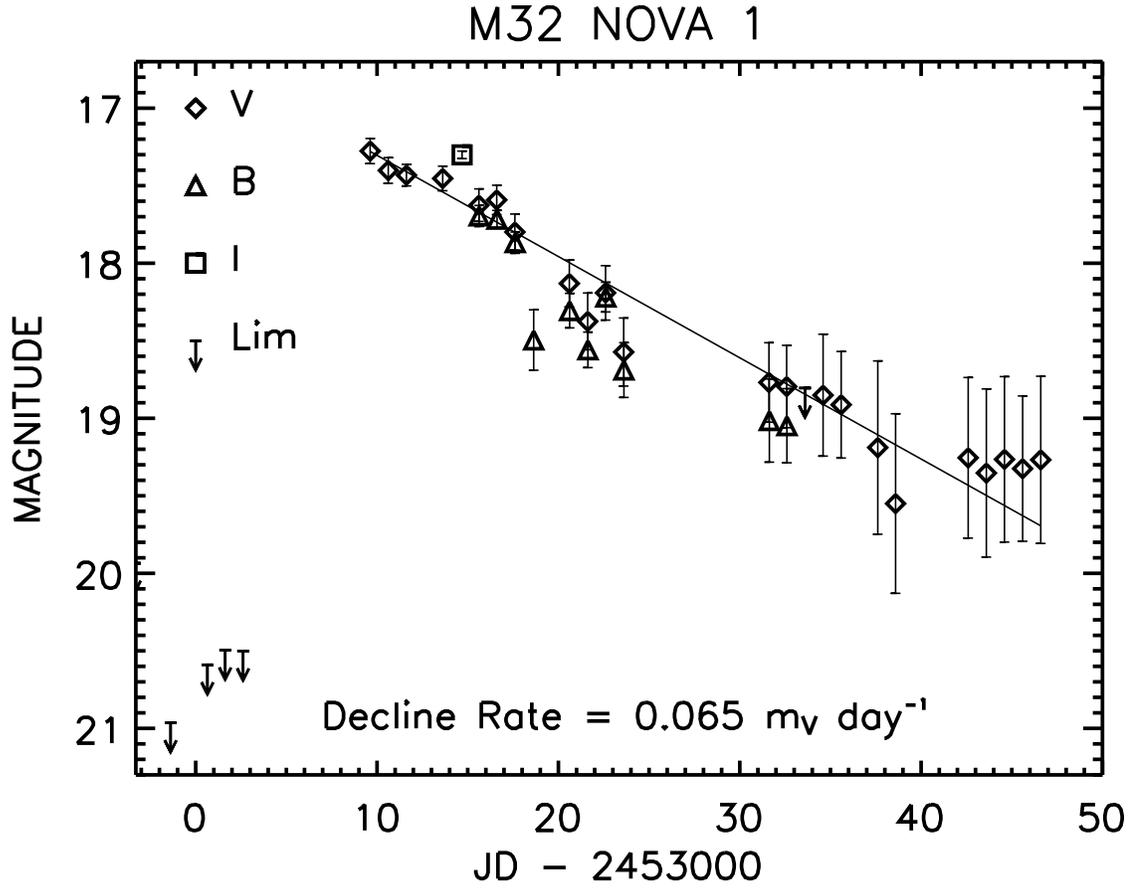}
\caption{Light curve of M32 nova 1 in V, B, and I-bands.  V-band points are
indicated by diamonds, B-band by triangles, and the I-band point by the square.
Frame limits are indicated by short horizontal lines with a downward pointing
arrow.  The decline rate of 0.065 $m_V$ day$^{-1}$ was determined from an 
error-weighted linear fit (shown by the thin line) to the V-band points.
\label{m32n1lc}
}
\end{figure}

\begin{figure}
\includegraphics[angle=90,scale=.70]{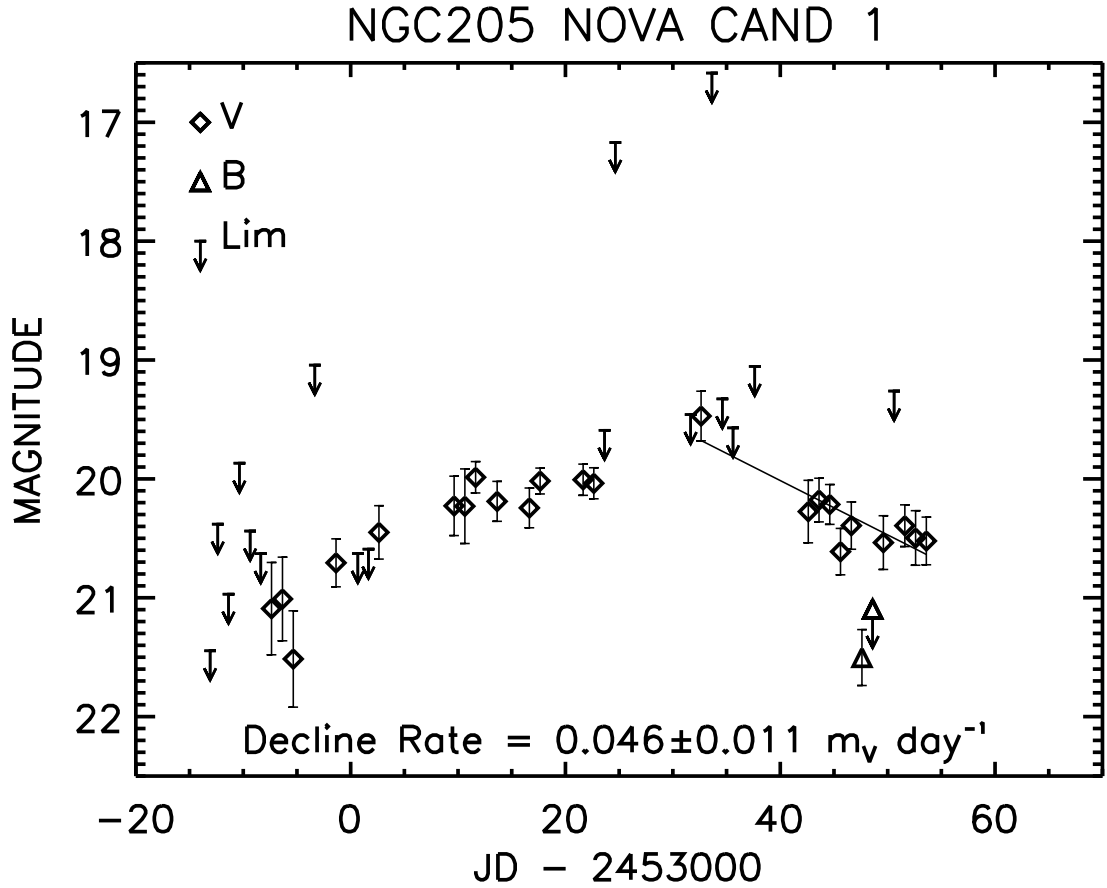}
\caption{Light curve of NGC 205 nova candidate 1 in V, and B-bands.  V-band points are
indicated by diamonds, and B-band by triangles.
Frame limits are indicated by short horizontal lines with a downward pointing
arrow.  The decline rate of 0.046 $m_V$ day$^{-1}$ was determined from an 
error-weighted linear fit (shown by the thin line) to the V-band points 
after maximum.
\label{n205n1lc}
}
\end{figure}

Figure~\ref{m32n1lc} presents the calibrated light curve for M32 nova 1,
and Figure~\ref{n205n1lc} presents the calibrated light curve for NGC 205
nova candidate 1.  The frame limits are plotted as short horizontal lines with
downward pointing arrows, while the open points with error bars are the
observations from Tables~\ref{m32n1_phot} and \ref{n205n1_phot}.

A simple linear fit was made to the decline portion of each light curve
in V to calculate the decline rate in $m_V$~day$^{-1}$.  The thin lines
in Figure~\ref{m32n1lc} and Figure~\ref{n205n1lc} show the resulting
fits.  Table~\ref{tab_nov_prop} presents the properties of the
light curves including the rise time and decline rate for each object.
The minimum magnitudes were determined from {\it HST} observations
(see \S\ref{verify}).

We calculated the average B$-$V color of the novae over the time
that they were observed in two colors and present the results in
Table~\ref{tab_nov_prop}, column 5.  For M32 nova 1, we used the 9 epochs
for which B and V were observed on the same night.  This gave $<$B$-$V$>$
= 0.14$\pm$0.08, which is typical of novae near maximum.  Because of
bad weather, and the faintness of NGC 205 nova candidate 1, we were unable 
to get
simultaneous B and V measurements.  The color we report was derived by
using the one B measurement on MJD 53047 and subtracting the average of
the two V points on MJD 53046 and MJD 53049.  The resulting color for
NGC 205 nova candidate 1 is redder than a typical nova ($<$B$-$V$>$ = 1.04$\pm$0.15),
but this could easily be due to short timescale fluctuations during the
decline phase.

\begin{figure}
\plotone{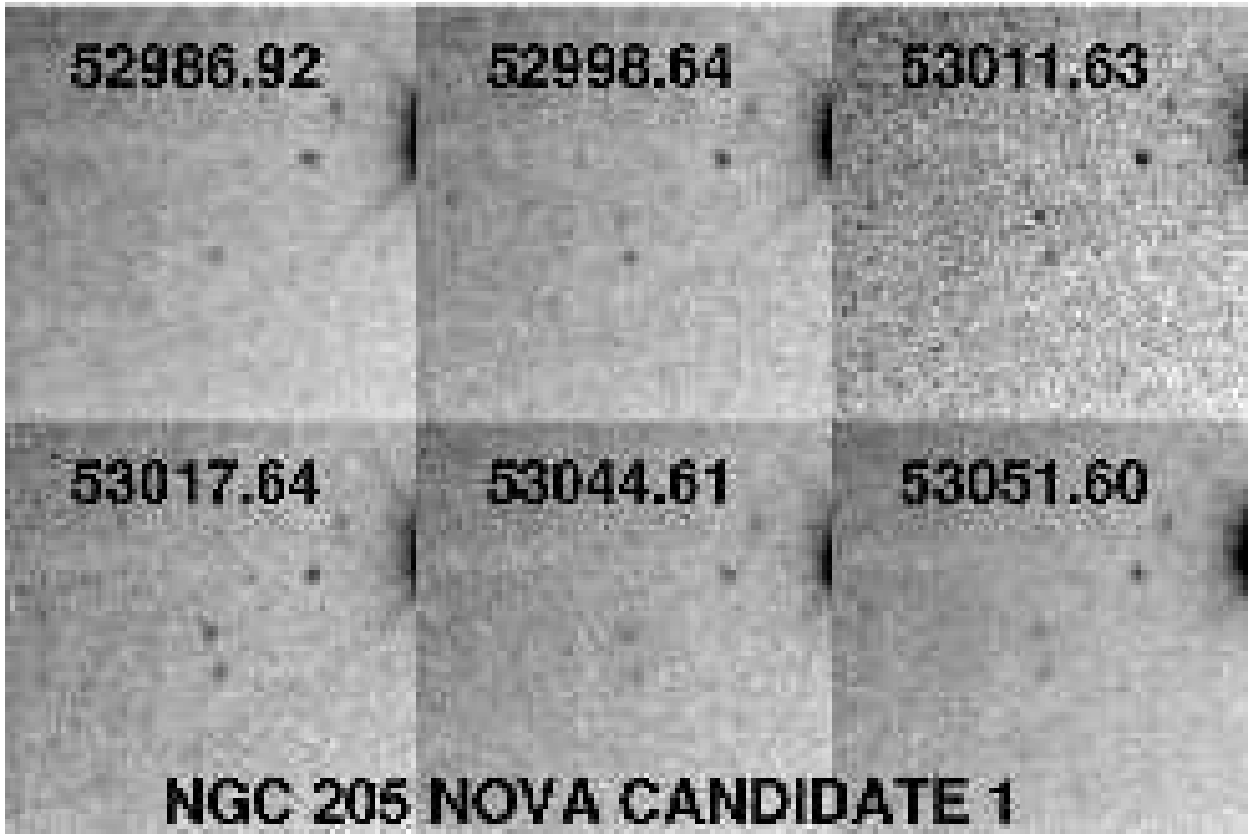}
\caption{Six zoomed V-band images of the nova candidate in NGC 205 spanning
the light curve in Figure~\ref{n205n1lc}.
Each image is 87 arcseconds on a side and has north at the top and east to
the left.  The nova candidate is centered in each image and the MJD of each
epoch is labeled above the candidate.
\label{n1ser}
}
\end{figure}

The light curve of the nova candidate in NGC 205 has a few unusual features.
It has
a long rise time, and it never reaches an intrinsic luminosity greater
than M$_V = -$5.1, which is low for a nova.  Bad weather produced less than
optimal coverage near the peak and we had to conclude our survey before the
candidate had completely faded.  Classical novae have been observed to
rise quite slowly (see, e.g., nova 29 in Arp 1956) and can have quite
inhomogeneous light curves \citep{arp56}.  If we missed the true peak of
the candidate and it was
only 0.5 mag higher this object would be in the range of the lowest luminosity
novae.  Also due to incomplete coverage of the decline portion, the decline
rate for this candidate is uncertain and could have been much slower.
To bolster the reality of this nova candidate we present a sample of zoomed 
V-band images in Figure~\ref{n1ser} that span the light curve.  Each image 
is labeled with the MJD and can be compared with photometry in 
Table~\ref{n205n1_phot}.

\section{Verifying the Novae\label{verify}}

In order to have confidence in our derived nova rates for these galaxies,
it is crucial that we be sure that the objects we discovered are indeed
novae and not some other kind of variable.  Spectroscopy is the best
way to verify a nova because of the telltale broad H emission lines.
Another way is to place a limit on the amplitude of the outburst.

\subsection{Verifying M32 Nova 1}

\begin{figure}
\includegraphics[angle=90,scale=.70]{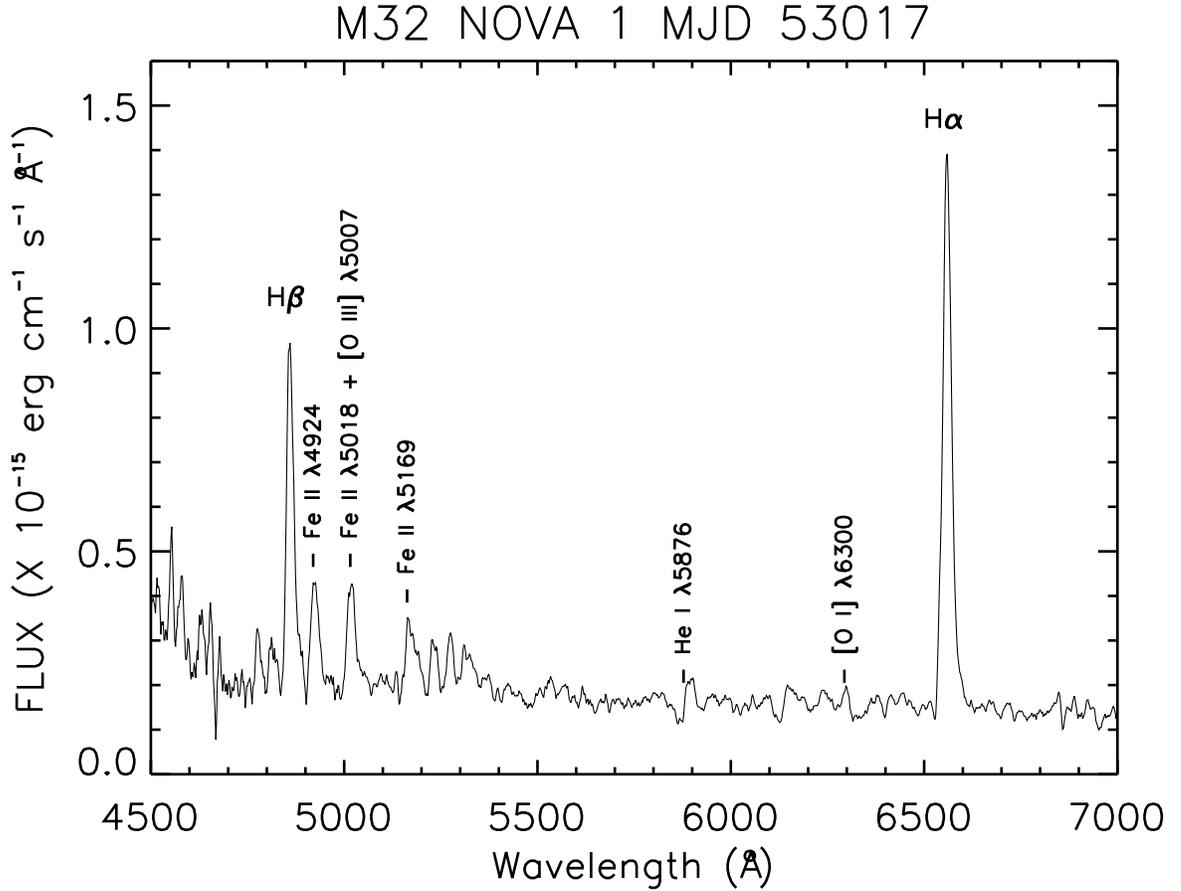}
\caption{Spectrum of M32 nova 1 taken on JD 2453017.60, 8 days after
discovery.  The broad H emission lines confirm it
as a classical nova.  The H$\alpha$ line has a half width at
half intensity of 640 km s$^{-1}$, at the low end of the velocity range 
(typically 300 to 3000 km~s$^{-1}$) for classical novae.  The strength of
the Fe II lines and the weakness or absence of the He and N lines is
consistent with the slow nova classification.  A spectrum taken on JD
2453022.60, 5 days later, is nearly identical to the spectrum presented
here, illustrating the slow spectral evolution of this nova.  Note that the 
He I line ($\lambda$5876) has a P Cygni profile and the H$\alpha$ line 
also shows a hint of absorption on the blue side.
\label{m32n1spec}
}
\end{figure}

Our spectral observations of M32 nova 1 confirm that it is indeed a nova.
They show broad H$\alpha$ and H$\beta$ in emission with a velocity
of expansion of 640 km s$^{-1}$, typical of a slow classical novae.
Figure~\ref{m32n1spec} shows the spectrum taken on JD 2453017.60, roughly
ten days after maximum.  Emission lines of Fe II can be seen just red-ward
of H$\beta$ which is also consistent with it being a typical slow nova.
A spectrum taken five days later is nearly identical, an indication of
the slow spectral evolution of this nova.

Because M32 is superposed on the outer disk of the larger M31, we also
wanted to confirm that this nova did indeed originate in M32.  We measured
the systemic heliocentric radial velocity of the nova from the H$\alpha$
lines and found it to have a velocity of $-$170 $\pm$ 6 km s$^{-1}$.
This is consistent with the radial velocity of M32 ($-$200 $\pm$ 6
km s$^{-1}$, Sandage \& Tammann 1981), if we account for the velocity
dispersion of M32 at a radius of 78".  The radial velocity of M31 is
$-$297 $\pm$ 1 km s$^{-1}$ \citep{san81}.  The part of M31's disk upon
which M32 is superposed is where most of the disk's rotation velocity would
be transverse to the line of sight and therefore could not make up the
difference in velocity between M31 and the nova.  This, combined with
the proximity of the nova to the nucleus of M32, argues against it
originating in M31.

In addition to these spectral confirmations, we also put a limit on
the nova outburst amplitude.  To do this, we used the {\it HST}\ \
WFPC2 observations in V (F555W) taken nearly a decade prior to outburst.
We also used the I (F814W) and F1042M images to verify that the progenitor
was not a bright red variable.  The nova position in all these filters
was in the WFPC2 chip WF3.

There were not enough stars in common with our Tenagra observations to
allow the precise determination of the position of the nova in the WF3
images directly.  We solved this by using the 4m 8k MOSAIC image from
the NOAO archive which went deep enough to pick up a significant number
of stars from the WF3 images.  First, we registered the MOSAIC image
to the WF3 image using 17 stars in common with the IRAF geomap task.
This produced a transformation fit with a root mean square (RMS) of 1.0
WF3 pixels.  We refined this transformation locally, in the region of
the nova, in the following way.  We registered the MOSAIC image to the
WF3 image using the above transformation.  We then extracted a small
subimage surrounding the position of the nova (75x87 WF3 pixels) in
the MOSAIC and the WF3 image.  The WF3 subimage was convolved with a
Gaussian to match the point spread function (PSF) of the MOSAIC image.
These two images were then cross correlated to detect any offset.
An offset in x of 0.35 px, and in y of -0.70 px, was detected with an
accuracy of 0.05 WF3 pixels, which improved the transformation locally
by at least a factor of ten.

We then produced a transformation from each of 14 Tenagra images (with the
nova well observed) to the MOSAIC image using over 400 stars in common.
This produced transformations with a typical RMS of less than 0.001
MOSAIC pixels.  We used the same cross-correlation technique described
above but were unable to detect any local offset in these transformations.
We transformed the nova positions to the MOSAIC coordinate system and
then used the MOSAIC to WF3 transformation, plus the detected offsets,
to put the nova positions in the WF3 coordinate system to better than 0.1
WF3 pixels.  We then used an error weighted average of the 14 positions
to get a final position in the WF3 images.  This position had an RMS of
0.04 WF3 pixels.

\begin{figure}
\plotone{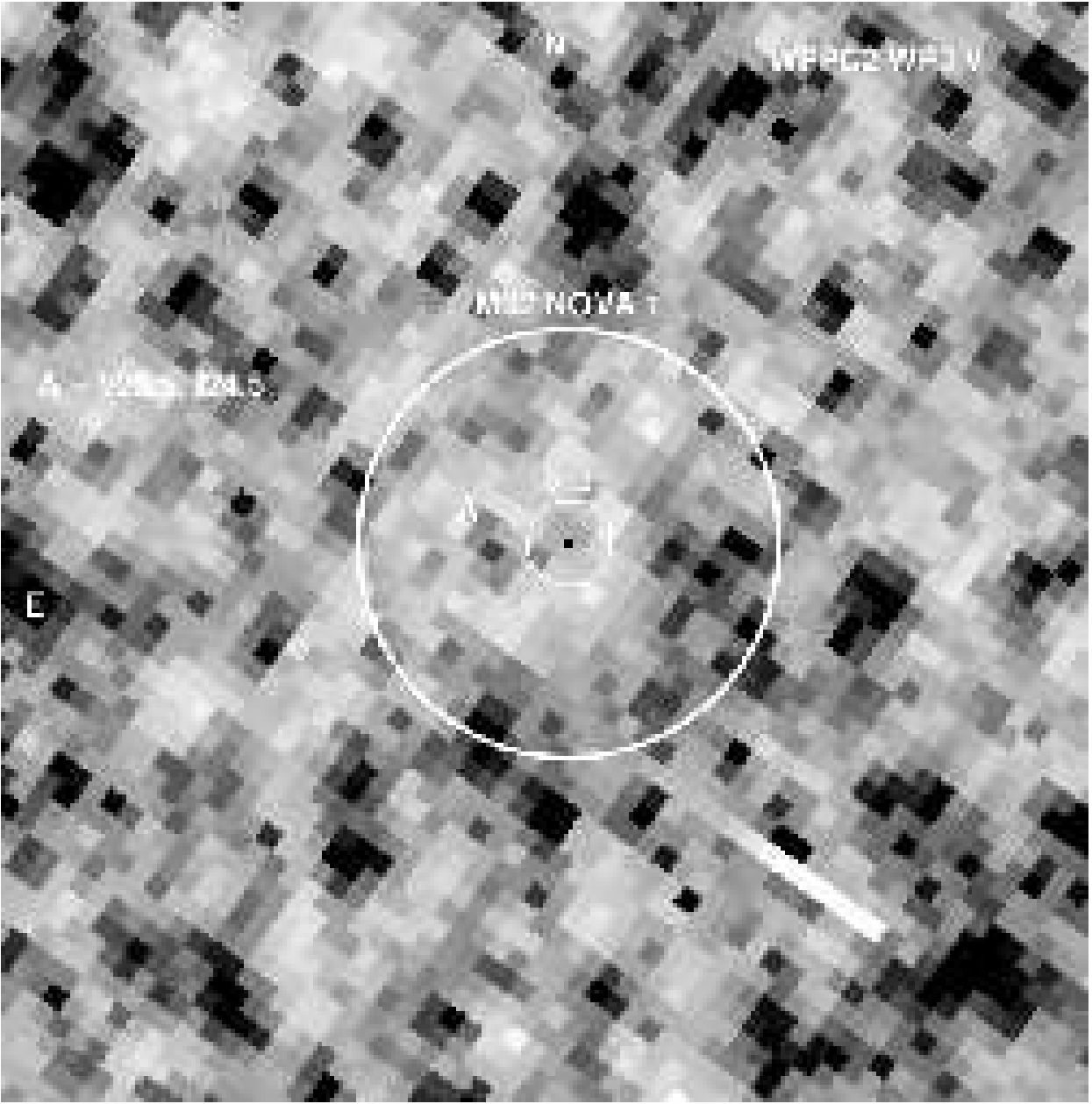}
\caption{{\it HST} WFPC2 WF3 V-band image of the region around M32 nova 1 taken
on JD 2449622.62.  North is up and East to the left.  The circles are
centered on the position of M32 nova 1.  The larger circle has
a radius of 1" and the smaller a radius of 0\arcpnt2.  The positional error
of M32 nova 1 is 0.1 WF pixels or 0\arcpnt01 and is represented as the
black dot at the center of the figure.  Star A has a V 
magnitude of 25.5, but is well outside the 1$\sigma$ error circle of M32 
nova 1.  The progenitor of M32 nova 1 is fainter than V of 26.0, implying 
an outburst amplitude of over 8.7 V magnitudes.
\label{m32n1hst}
}
\end{figure}

Figure~\ref{m32n1hst} shows the region in the WF3 V image surrounding
the nova, with a field of view of 3 arcseconds on a side.  We do not
convincingly detect the progenitor of M32 nova 1.  The star indicated
by the letter A is at V of 25.5 and I of 24.5, but is well outside
the position error circle defined by the RMS of the nova positions.
We estimate that the progenitor of this nova was fainter than V of 26.0
at the epoch that these data were taken.  This gives an amplitude of at
least 8.7 magnitudes in V for the nova outburst.  The WFPC2 I and F1042M
images were well registered with the V image and nothing was detected
in them at the position of the nova.  We conclude, therefore, that M32
nova 1 could not be a bright red variable and is, indeed, a nova.

\subsection{Verifying NGC 205 Nova Candidate 1}

This variable presented more of a challenge.  Its faintness precluded spectral
observations with the MDM 2.4 m.  In this case, the {\it HST} observations
were examined in an attempt to bolster the nova classification by
constraining the outburst amplitude.  We were fortunate that
the ACS WFC observations of this area included 19 to 21 stars that
could be used to directly register the Tenagra images.  We chose 5 of
these images, in which the candidate was well observed, to register to the
ACS image.  Before centroiding objects in the ACS image, we convolved
it with a Gaussian of appropriate width to bring the ACS
resolution down to the resolution of the Tenagra images.  The IRAF task
geomap was used to calculate the transformations using a 2nd order
polynomial including cross terms.  The average RMS for the 5 
transformations was 2 ACS pixels in x and y.
We used these transformations to place the candidate in the ACS WFC image.
We then calculated the error-weighted average position for the candidate ACS
position which had an RMS of 6.4 ACS pixels.  To check for local offsets
we used two stars within 25" of the candidate that were visible in both the
Tenagra and the ACS image.  We checked the transformed positions of
these stars and computed their error weighted average position, which
had an RMS of 2.0 ACS pixels.  These average positions showed no
local offset with respect to the ACS to 0.5 ACS pixels.  The larger scatter
in the candidate positions is due to the faintness of the candidate compared 
with the stars we used for registration.

\begin{figure}
\plotone{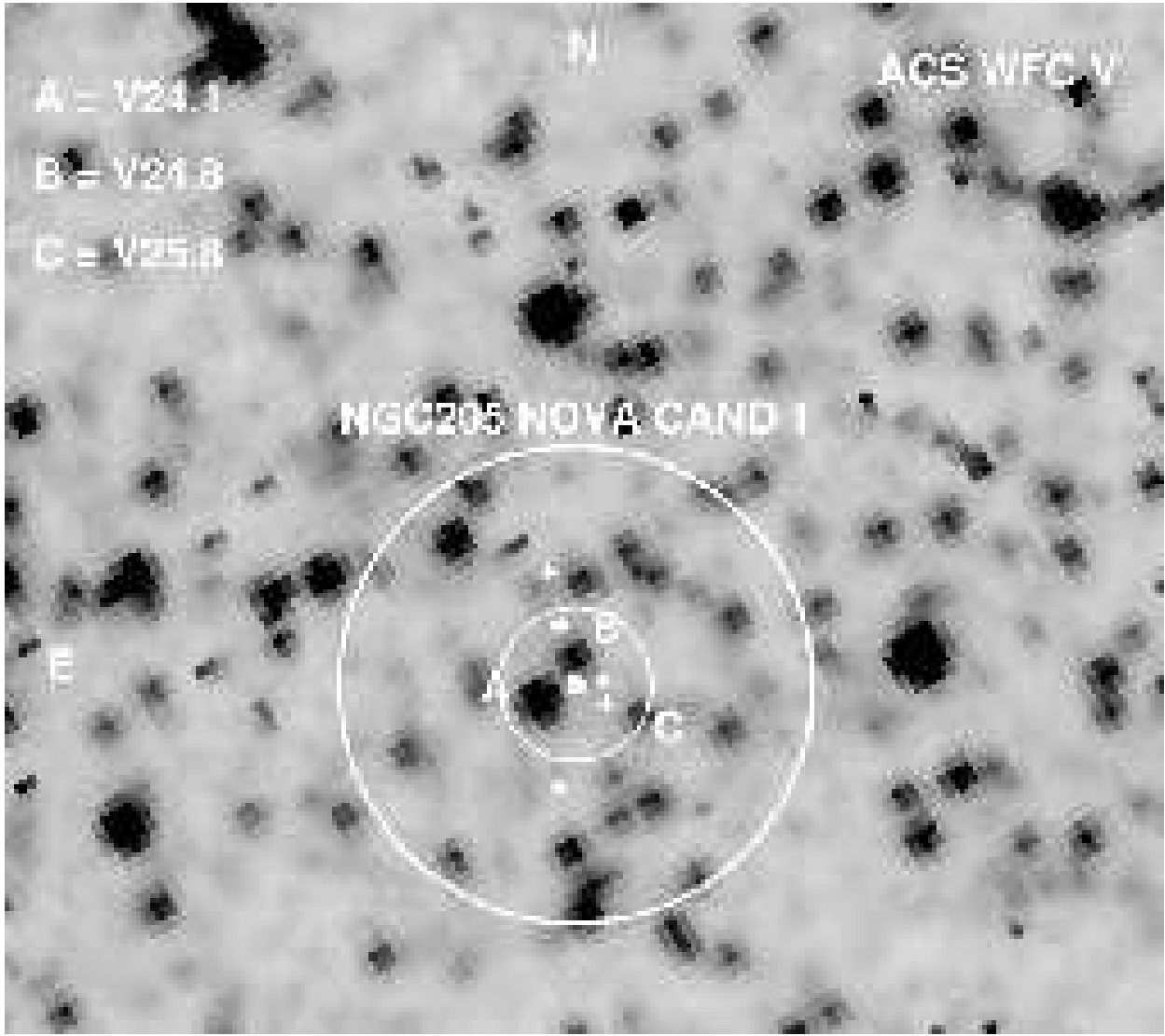}
\caption{{\it HST} ACS WFC V-band image of the region around NGC 205 
nova candidate 1 taken on JD 2452525.83.  North is up and East to the left.  
The circles are centered on the error-weighted centroid of the 5
transformed positions for the candidate.  The larger circle has
a radius of 1" and the smaller a radius of 0\arcpnt32, the RMS error
of the positions of NGC 205 nova candidate 1.  Star A has a V magnitude of 24.1, 
B a V magnitude of 24.8, and C a V magnitude of 25.8.
The nova progenitor could conceivably be anywhere within the error circle, 
so we use the brightest star within the circle to place a lower limit on
the outburst amplitude of the nova candidate of $>$ 4.6 Vmag.
\label{n205n1hst}
}
\end{figure}

Figure~\ref{n205n1hst} shows the position of the candidate nova in the ACS WFC V
(F606W) image, with a 5 arcsecond field of view.  The crosses mark the five
transformed candidate positions and the smallest circle marks the error-weighted
centroid of these positions. The next larger circle shows
the RMS error circle for the nova candidate positions and the largest circle is 1
arcsecond in radius.  The three stars within the RMS error circle have
V magnitudes of 24.1, 24.8, and 25.8.  The progenitor of the outbursting
object we observed could conceivably be anywhere within the RMS error
circle and therefore we place a lower limit on the outburst
amplitude using the brightest of the three stars.  This gives a limit of
$>$4.6 Vmag for the outburst amplitude of the nova candidate.  If star A were in fact
the variable that produced our nova candidate detection it would have an absolute V
magnitude in quiescence of $M_V = -0.4\pm0.2$ at the distance of NGC 205.

There are few variables with amplitudes of 4.6 magnitudes or greater, but
we must consider each in turn.  The shape of the light curve eliminates
background supernovae since the rise appears slower than the decline.  The
peak magnitude of $M_V = -5.1$ is too high for a Mira-type variable in NGC
205 and is too low for a Hubble-Sandage variable in NGC 205.  The light
curve also eliminates a foreground dwarf nova since these exhibit rise
rates much higher than we observe.

A more likely alternative to the nova classification is a microlensing
event.  The flattening of the light curve after the initial rise starting
at JD 2453010 would argue against this classification, however the errors
are too large and the coverage near the peak is not good enough to
be sure.  We must resort to a statistical argument to bolster the claim
that this was not a lensing event.

In a recent paper reporting the results of a microlensing survey of M31,
\citet{ugl04} found 4 genuine events in 200 epochs over a 3 year campaign 
covering 560 arcmin$^2$ with 1.3 m and 1.8 m telescopes. 
\citet{dej04} report finding 14 microlensing candidate events in M31 
in 100 epochs over a 2
year study covering 0.57 square degrees with a 2.5 m telescope.  These
studies were concentrated near the center of M31 and comprised a total of
over 300 epochs.  Our study covered 196 arcmin$^2$ with a 0.8 m telescope
and covered 90 epochs.  Ignoring the effects of telescope size and the number
of lenses and targets in the survey areas, we can divide out the areal
coverage and number of epochs and compute that we would see about 1 event in
our survey.  A more realistic measure of the microlensing rate in our survey 
would have to account for the shallower frame limits provided by a smaller
telescope and the much lower number of M31 halo lenses due to the distance 
from M31 (36.5 arcmin) and the much lower mass of NGC 205.  These factors 
reduce the microlensing rate in our survey by over two orders of magnitude 
and make it very unlikely that the variable in NGC 205 is a microlensing 
event.

While the classification of the variable in NGC 205 is not ironclad, it
appears that a classical nova is the most likely one.  Because of the
uncertainty, we will compute nova rates both with and without the nova 
candidate in the following sections.

\section{The Nova Rate}

A raw bulk nova rate for each galaxy can be obtained
simply by dividing the observed number of novae that erupted during
the survey by the time covered.  This gives $R = 1 / 0.355$yr $= 2.82$
yr$^{-1}$ for M32 and $R = 1 / 0.375$yr $= 2.67$ yr$^{-1}$ for NGC 205
(this becomes an upper limit without the nova candidate).
For NGC 147 and NGC 185 we can place a limit on the nova rate by saying
that it is no greater than the inverse of the survey time.  For NGC 147
this gives $R < 1 / 0.315$yr $= 3.18$ yr$^{-1}$ and for NGC 185 this
gives $R < 1 / 0.323$yr $= 3.10$ yr$^{-1}$.

\subsection{The Monte Carlo Approach}

\citet{sha01} describe a Monte Carlo technique which uses the maximum
magnitudes and decline rates of novae and their survey faint limit to
find the most probable nova rate in their survey region.  We used the
V-band maximum magnitudes and decline rates reported in \citet{arp56},
and \citet{ros73}, combined with our individual epoch frame limits,
to perform a similar Monte Carlo experiment to derive nova rates for
each galaxy we observed.

This technique makes many independent estimates of the observed nova rate
in the given galaxy as a function of the true nova rate $[N_{obs}(N_t)]$.
For a given trial estimate of $N_t$, the true rate, we choose a random
set of novae (which specify the maximum magnitudes and decline rates of
real novae) and outburst times and use the frame limits to calculate the
number of observed novae, using the candidate criteria described above.
We repeat this 10$^5$ times and record how many times we recover
the number of nova candidates actually observed in that galaxy.
The estimate of the true nova rate $N_t$ is then incremented and the
process is repeated.  This produces a probability distribution for
$N_t$ in the given galaxy.  The best estimate for $N_t$ is that which
corresponds to the peak of this distribution.

\begin{figure}
\includegraphics[angle=90,scale=.70]{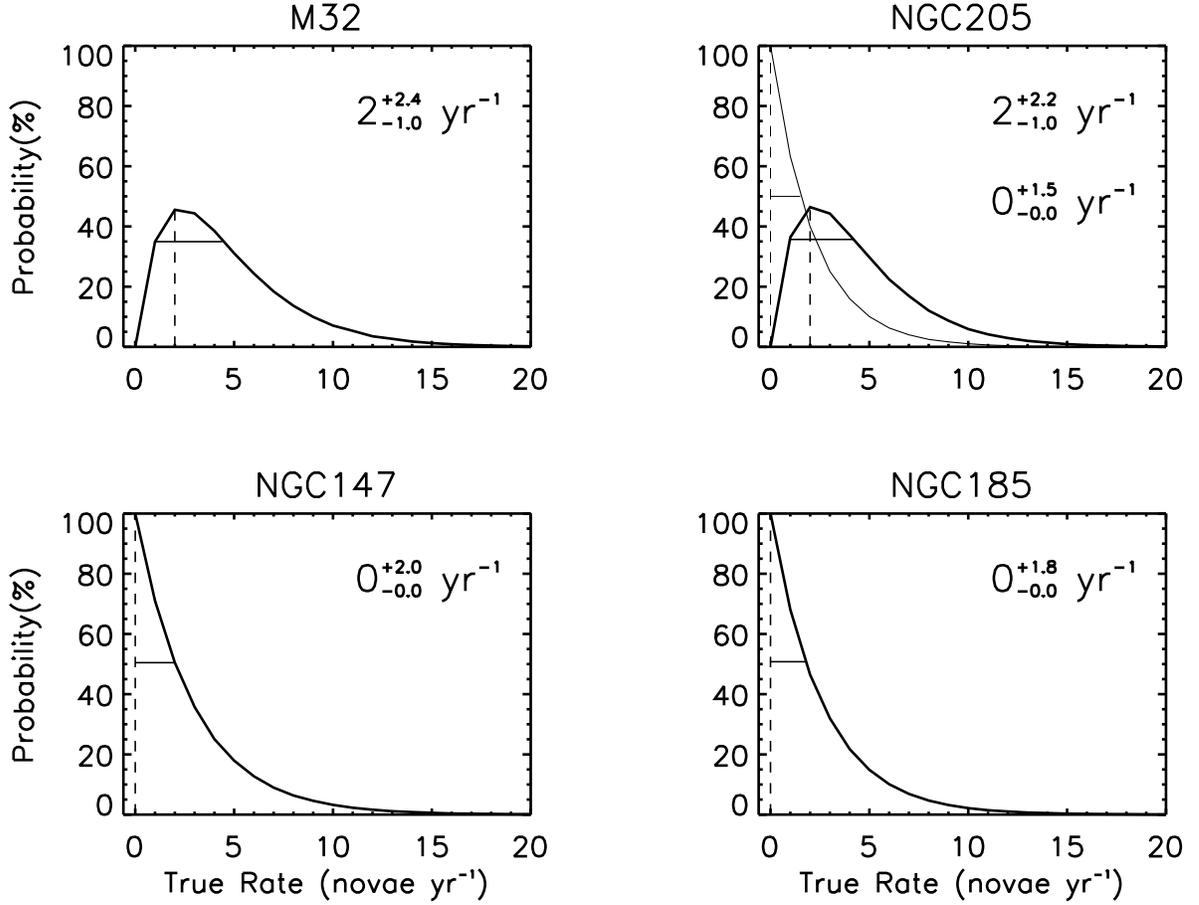}
\caption{Probability distributions from Monte Carlo simulations of the true
nova rates in M32, NGC 205, NGC 147, and NGC 185, based on individual epoch
frame limits, well observed V-band nova maximum magnitudes and decline
rates \citep{arp56,ros73}, and observed numbers of novae.  The observed 
number of novae used for the
simulations for M32 and NGC 205 were one each, while the observed number of
novae for the NGC 147 and NGC 185 simulations were zero each.
For NGC 205, the probability distribution for zero observed
novae is also plotted as the thinner solid line.  The horizontal lines show what
part of each distribution around the peak encloses half the probability and 
define the error limits on the bulk rates.
\label{mc}
}
\end{figure}

Figure~\ref{mc} shows the probability distributions for each galaxy.
The shapes of the probability distributions depend on the number of novae 
observed in each galaxy and the temporal distribution and depths of the 
galaxy's survey epochs.  The range encompassing half of the probability
distribution surrounding the peak is indicated by the solid horizontal 
lines and defines the error limits for the bulk nova rates reported in 
the figure.  For NGC 205, we overplotted the 'no nova' probability 
distribution as the thinner solid line.

Column 2 of Table~\ref{tab_nova_rate} presents the results of the
simulations for each of the galaxies.  A total value for all the local
group dwarf ellipticals surveyed is presented at the bottom of the table
both with and without the nova candidate in NGC 205.

\subsection{The Luminosity Specific Nova Rate}

To facilitate the comparison of nova rates across a broad range of galaxy
luminosities a suitable normalization must be found.  The infrared is
used as a measure of stellar mass to avoid large fluctuations due to a
few bright blue stars.  The 2MASS offers a consistent photometric system
for this normalization, however \citet{fer03} found discrepancies between
the K-band magnitudes from the 2MASS Large Galaxy Atlas \citep{jar03} and
those derived from integrated optical magnitudes and optical to infrared
colors.  We chose to use the integrated, extinction corrected V-band
magnitudes from \citet{dev91} and (V$-$K)$_0$ colors from \citet{aar78}
and \citet{fro78} to perform our normalization.

Table~\ref{tab_nova_rate} lists the relevant data and references for
M81 and the four dwarf ellipticals of this study.  We included the
2MASS K-band magnitudes, after correcting for galactic extinction, to
compare with the K-band magnitudes derived from the optical magnitudes
and colors.  The difference is $+$0.36 mag for M32 and $-$0.29 mag
for NGC 205, illustrating the discrepancy.  No (V$-$K)$_0$ colors were
available for NGC 147 and NGC 185, so we adjusted K$_{2MASS,0}$ by the
0.2 magnitude systematic offset between the two systems found by 
\citet{fer03}.  To arrive at the LSNR, the bulk nova rates are divided 
by the K-band luminosities, expressed in 10$^{10} L_{\odot,K}$.

\begin{figure}
\includegraphics[angle=90,scale=.70]{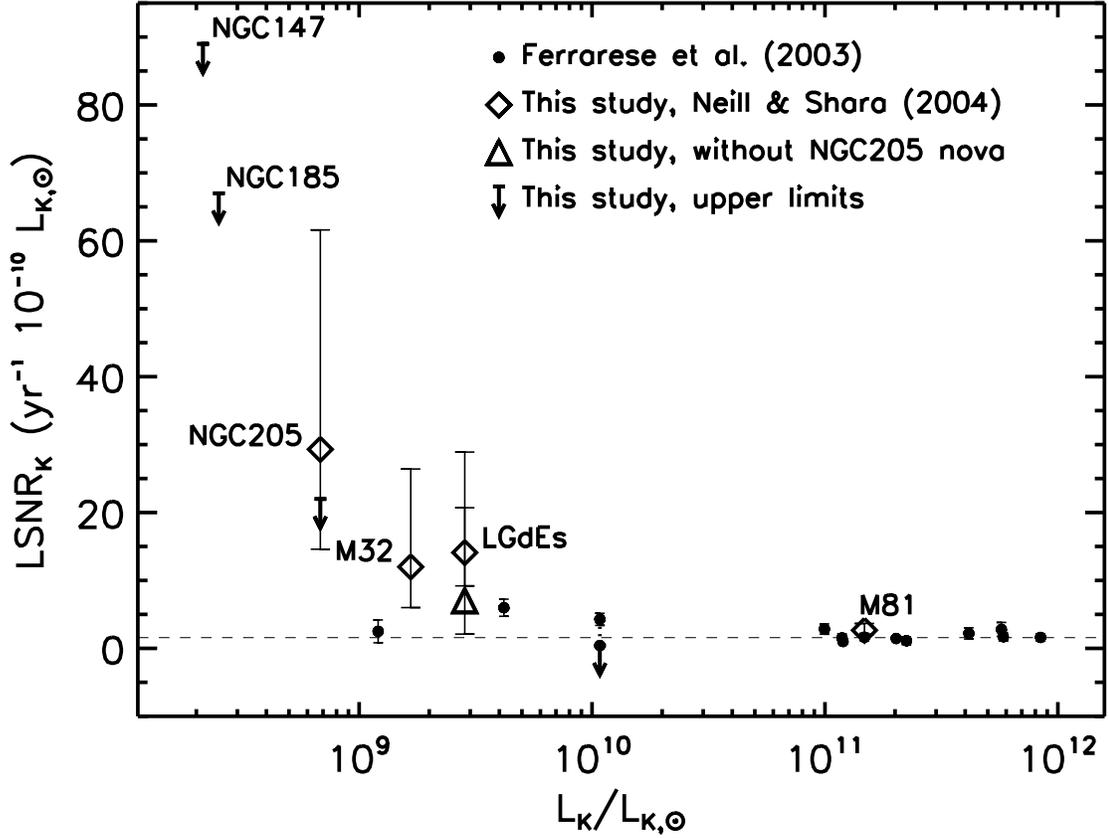}
\caption{LSNR versus K-band luminosity in Solar units.  
The small filled circles are from data in \citet{fer03}.  As in their 
Figure~18, the dotted line connects the two (discrepant) values for M33.
The large open diamonds are the results from \citet{nei04} and
Table~\ref{tab_nova_rate} of this work.  The short horizontal lines with
the downward pointing arrows are the upper limits from
Table~\ref{tab_nova_rate} for NGC 147, NGC 185, and
NGC 205 (assuming no nova) and the open triangle is the total rate for the
four dEs without the nova candidate in NGC 205.  The dashed line is the constant 
LSNR of 1.58 yr$^{-1} [10^{10} L_{\odot,K}]^{-1}$ derived in \citet{fer03}.
\label{lsnr}
}
\end{figure}

We plot our results, and the result for M81 from \citet{nei04}, as
open diamonds in Figure~\ref{lsnr} along with data from \citet{fer03},
Table~5, plotted as small filled circles.  The two discrepant points
for M33 are connected with a dotted line.  The upper limits for NGC 147,
NGC 185, and NGC 205 are plotted as short horizontal lines with a downward
pointing arrow.  The open triangle is the point from this study for 
all the dwarf ellipticals assuming the nova candidate in NGC 205 was misclassified.
The horizontal dashed line is the average luminosity specific nova rate 
(LSNR) from \citet{fer03}.

The LSNR for all the dEs that includes the nova candidate in NGC 205 of
14.1$^{+14.8}_{-4.9}$ yr$^{-1} [10^{10} L_{\odot,K}]^{-1}$ is 2.5$\sigma$
higher than the constant LSNR derived in \citet{fer03} of 1.58$\pm$0.16
yr$^{-1} [10^{10} L_{\odot,K}]^{-1}$.  The LSNR  without the
nova candidate in NGC 205 of 7.0$^{+13.7}_{-4.9}$ yr$^{-1}[10^{10} L_{\odot,K}]^{-1}$
is only 1.1$\sigma$ higher.  These numbers are
suggestive of a possible increase in the nova rate for lower mass systems,
but at this point no conclusions can be drawn.  If a higher rate for the
Local Group dEs is verified in future surveys there are two possible
explanations.  Either the interacting binary fraction is higher for these
systems than for higher mass systems, or the completeness is higher due to
their proximity and lack of dust.

\section{Future Work}

It is important to improve the statistics for these low mass, nearby
systems.  We will continue to monitor M32, NGC 205, NGC 147, and NGC 185
for several years.  The nova candidate we discovered in NGC 205 would be
the lowest luminosity nova ever seen, if verified.  It is possible 
this nova candidate
represents a population of low luminosity novae with M$_V \sim$ -5.0.  The 
implication is that faint novae are being missed in surveys of more distant, 
larger, and dustier systems producing a systematic underestimate of the nova 
rates in these systems.  It is crucial to continue to survey nearby systems
where we can more easily detect these low luminosity novae and to acquire a
spectrum during outburst of one of them to verify the nova classification.

Eliminating systematic errors in bulk nova rates is essential to using
the LSNR vs. luminosity diagram to determine the factors that influence
the formation and evolution of the close, interacting binaries that
produce novae.  With large-format cameras and the availability of
service telescope time on a nightly basis, we can determine precise
nova rates and distributions for these nearby small systems.  We must
also complement this with the best rates and distributions possible for
larger, more distant systems.  If the low luminosity nova candidate we discovered
in NGC 205 represents a new nova population, we must be sure to sample
this population in nova surveys of the larger, more distant systems.
This will be observationally expensive.

\section{Conclusions}

1. Using the Monte Carlo technique we derive a bulk nova rate for M32 of
2$^{+2.4}_{-1.0}$ yr$^{-1}$ and for NGC 205 of 2$^{+2.2}_{-1.0}$ yr$^{-1}$.
We could not verify the nova candidate in NGC 205 conclusively and so we also report 
upper limits
on the bulk nova rates based on Monte Carlo simulations for NGC 205 of 
1.5 yr$^{-1}$, for NGC 147 of 2 yr$^{-1}$, and for NGC 185 of 1.8
yr$^{-1}$.  We also calculated a combined bulk nova rate for the four dEs
of 4$^{+4.2}_{-1.4}$ yr$^{-1}$ (2$^{+3.9}_{-1.4}$ yr$^{-1}$ without the NGC
205 nova candidate).

2. Using V$_{T,0}$ magnitudes and (V$-$K)$_0$ colors to derive K-band
luminosities to normalize these rates produces an LSNR for M32 of
12.0$^{+14.4}_{-6.0}$ yr$^{-1} [10^{10} L_{\odot,K}]^{-1}$, and for
NGC 205 of 29.3$^{+32.3}_{-14.7}$ yr$^{-1} [10^{10}L_{\odot,K}]^{-1}$.
We also report upper limits on the LSNR for NGC 205 of 22.0 yr$^{-1}
[10^{10}L_{\odot,K}]^{-1}$ without the nova candidate, for NGC 147 of 89.0 yr$^{-1}
[10^{10}L_{\odot,K}]^{-1}$, and for NGC 185 of 67.0 yr$^{-1}
[10^{10}L_{\odot,K}]^{-1}$.  Using the combined bulk rate and the combined
K-band luminosity of the four dEs we report an LSNR of 14.1$^{+14.8}_{-4.9}$
yr$^{-1} [10^{10} L_{\odot,K}]^{-1}$ for the total
(7.0$^{+13.7}_{-4.9}$ yr$^{-1}[10^{10} L_{\odot,K}]^{-1}$ without the NGC 
205 nova candidate).  These total LSNRs are marginally higher than predicted by 
extrapolating a constant LSNR determined at high luminosities to low 
luminosities.



\acknowledgments

We are very grateful to John Thorstensen for obtaining the spectra of
M32 nova 1, allowing us to confirm its nature. We are also thankful
to him for obtaining and astrometrically calibrating the I-band epoch
of this nova, allowing us to refine its position.  We acknowledge the
valuable assistance of the Tenagra Observatory team, Michael Schwartz,
and Paulo Holvorcem, in determining coordinates for the novae and in
preparing the IAU circulars announcing their discovery.  We also thank the
anonymous referee for valuable comments on the manuscript.

This research has made use of the NASA/ IPAC Infrared Science Archive,
which is operated by the Jet Propulsion Laboratory, California Institute
of Technology, under contract with the National Aeronautics and Space
Administration.  This publication makes use of data products from the
Two Micron All Sky Survey, which is a joint project of the University of
Massachusetts and the Infrared Processing and Analysis Center/California
Institute of Technology, funded by the National Aeronautics and Space
Administration and the National Science Foundation.

This research draws upon data provided by Dr. Philip Massey as distributed
by the NOAO Science Archive. NOAO is operated by the Association of
Universities for Research in Astronomy (AURA), Inc. under a cooperative
agreement with the National Science Foundation.




\clearpage


\clearpage

\clearpage

\clearpage

\clearpage

\clearpage

\clearpage

\clearpage

\clearpage

\clearpage

\clearpage



\begin{deluxetable}{lcrrlccc}
\tabletypesize{\footnotesize}
\tablecolumns{8}
\tablecaption{Observations\label{tab_obs}}
\tablewidth{0pt}
\tablehead{
\colhead{}	 & \colhead{}	    & \colhead{Epochs} & \colhead{Exp.} &
\colhead{Start} & \colhead{End} & \colhead{Span}   & \colhead{Novae} \\
\colhead{Galaxy} & \colhead{Filter} & \colhead{(N)}    & \colhead{(s)}  &
\colhead{(MJD)}	& \colhead{(MJD)} & \colhead{(days)} & \colhead{(N)}}

\startdata
\cutinhead{Tenagra 0.8m Observations}
M32	& V	& 90	&900 & 52916.83	& 53046.60	& 129.8		& 1\\
	& B	& 14	&900 & 53015.63	& 53037.64	& \nodata	&\\
\\
NGC 205	& V	& 89	&900 & 52916.83	& 53053.60	& 136.8		& 1\\
	& B	& 2	&900 & 53047.61	& 53048.60	& \nodata	&\\
\\
NGC 147	& V	& 78	&900 & 52927.73	& 53042.62	& 114.9		& 0\\
\\
NGC 185	& V	& 84	&900 & 52927.79	& 53045.62	& 117.8		& 0\\
\cutinhead{MDM 2.4m Observations}
M32	& I	& 1	&60  & 53014.70	& \nodata	& \nodata	&1\\
	& Spec	& 2	&960 & 53017.6	& 53022.6	& 5.0		&1\\
\cutinhead{{\it HST} WFPC2 Observations}
M32	& F555W	& 1	&1600 & 49622.9	& \nodata	& \nodata	&0\\
	& F814W & 1	&1200 & 49622.9	& \nodata	& \nodata	&0\\ 
	& F1042M& 1	&3000 & 49622.9 & \nodata	& \nodata	&0\\
\cutinhead{KPNO 4m Observations}
M32	& V	& 1	&2400 & 52528.3 & \nodata	& \nodata	&0\\
\cutinhead{{\it HST} ACS Observations}
NGC 205	& F606W	& 1	&497  & 52525.8	& \nodata	& \nodata	&0\\
\enddata
\end{deluxetable}

\clearpage
\begin{deluxetable}{llccrr}
\tablecaption{Nova Positions\label{tab_nov_pos}}
\tablewidth{0pt}
\tablehead{
\colhead{} & \colhead{} & \multicolumn{2}{c}{Position} & \colhead{} & 
	\colhead{}\\
\colhead{} & \colhead{} & \multicolumn{2}{c}{(J2000)} & \colhead{} & 
	\colhead{Nuclear Distance}\\
\cline{3-4} \\
\colhead{Galaxy} & \colhead{Object} & \colhead{RA} & \colhead{Dec} &
\colhead{Detections} & \colhead{(arcsec)}}

\startdata
M32 	& Nova 1	& 00 42 44.991 & +40 53 04.76	& 33	& 78	\\
NGC 205	& Nova Cand 1	& 00 40 15.216 & +41 37 29.68	& 24 	& 230	\\
\enddata
\end{deluxetable}

\clearpage

\begin{deluxetable}{rrcccrccrcc}
\tabletypesize{\scriptsize}
\tablecaption{M32 Nova 1 Magnitudes\label{m32n1_phot}}
\tablewidth{0pt}
\tablehead{
 & & & & \colhead{Fr. Limit} & & & \\
\colhead{Nova} & 
\colhead{MJD} & \colhead{$m_V$} & \colhead{Err($m_V$)} & \colhead{$m_V$} &
\colhead{MJD} & \colhead{$m_B$} & \colhead{Err($m_B$)} &
\colhead{MJD} & \colhead{$m_I$} & \colhead{Err($m_I$)}
}

\startdata

M32 1 & 52998.62 & \nodata & \nodata & 20.96 & \nodata & \nodata & \nodata 
& \nodata & \nodata & \nodata \\ 
 & 53000.65 & \nodata & \nodata & 20.59 & \nodata & \nodata & \nodata 
 & \nodata & \nodata & \nodata \\ 
 & 53001.63 & \nodata & \nodata & 20.50 & \nodata & \nodata & \nodata
 & \nodata & \nodata & \nodata \\ 
 & 53002.61 & \nodata & \nodata & 20.50 & \nodata & \nodata & \nodata
 & \nodata & \nodata & \nodata \\ 
 & 53009.62 & 17.28 &  0.08 & 20.21 & \nodata  & \nodata & \nodata
 & \nodata & \nodata & \nodata \\ 
 & 53010.62 & 17.40 &  0.08 & 19.28 & \nodata  & \nodata & \nodata
 & \nodata & \nodata & \nodata \\ 
 & 53011.62 & 17.43 &  0.07 & 20.54 & \nodata  & \nodata & \nodata
 & \nodata & \nodata & \nodata \\ 
 & 53013.62 & 17.45 &  0.08 & 20.63 & \nodata  & \nodata & \nodata
 & \nodata & \nodata & \nodata \\ 
 &\nodata &\nodata &\nodata &\nodata &\nodata &\nodata &\nodata
 & 53014.70 & 17.30 & 0.02 \\
 & 53015.62 & 17.63 &  0.10 & 19.90 & 53015.63 & 17.70 &  0.07
 & \nodata & \nodata & \nodata \\ 
 & 53016.60 & 17.59 &  0.09 & 20.85 & 53016.62 & 17.72 &  0.06
 & \nodata & \nodata & \nodata \\ 
 & 53017.60 & 17.80 &  0.12 & 20.97 & 53017.62 & 17.87 &  0.07
 & \nodata & \nodata & \nodata \\ 
 & 53018.62 & \nodata & \nodata & 19.24 & 53018.64 & 18.50 &  0.19
 & \nodata & \nodata & \nodata \\ 
 & 53020.60 & 18.13 &  0.15 & 20.44 & 53020.62 & 18.31 &  0.11
 & \nodata & \nodata & \nodata \\ 
 & 53021.61 & 18.38 &  0.18 & 20.72 & 53021.63 & 18.56 &  0.11
 & \nodata & \nodata & \nodata \\ 
 & 53022.60 & 18.19 &  0.18 & 20.88 & 53022.61 & 18.22 &  0.10
 & \nodata & \nodata & \nodata \\ 
 & 53023.60 & 18.57 &  0.22 & 19.74 & 53023.61 & 18.69 &  0.18
 & \nodata & \nodata & \nodata \\ 
 & 53031.61 & 18.77 &  0.26 & 19.84 & 53031.63 & 19.02 &  0.27
 & \nodata & \nodata & \nodata \\ 
 & 53032.59 & 18.80 &  0.27 & 19.85 & 53032.60 & 19.05 &  0.24
 & \nodata & \nodata & \nodata \\ 
 & 53033.60 & \nodata & \nodata & 18.80 & \nodata & \nodata & \nodata
 & \nodata & \nodata & \nodata \\ 
 & 53034.59 & 18.85 &  0.39 & 19.34 & \nodata & \nodata & \nodata
 & \nodata & \nodata & \nodata \\ 
 & 53035.59 & 18.91 &  0.34 & 19.61 & \nodata & \nodata & \nodata
 & \nodata & \nodata & \nodata \\ 
 & 53037.61 & 19.19 &  0.56 & 19.12 & \nodata & \nodata & \nodata
 & \nodata & \nodata & \nodata \\ 
 & 53038.59 & 19.55 &  0.58 & 19.44 & \nodata & \nodata & \nodata
 & \nodata & \nodata & \nodata \\ 
 & 53042.60 & 19.25 &  0.52 & 19.66 & \nodata & \nodata & \nodata
 & \nodata & \nodata & \nodata \\ 
 & 53043.60 & 19.35 &  0.54 & 20.74 & \nodata & \nodata & \nodata
 & \nodata & \nodata & \nodata \\ 
 & 53044.60 & 19.27 &  0.53 & 20.71 & \nodata & \nodata & \nodata
 & \nodata & \nodata & \nodata \\ 
 & 53045.60 & 19.32 &  0.47 & 20.86 & \nodata & \nodata & \nodata
 & \nodata & \nodata & \nodata \\ 
 & 53046.60 & 19.27 &  0.54 & 20.81 & \nodata & \nodata & \nodata
 & \nodata & \nodata & \nodata
\enddata
\end{deluxetable}

\begin{deluxetable}{rrcccrcc}
\tabletypesize{\scriptsize}
\tablecaption{NGC 205 Nova Candidate 1 Magnitudes\label{n205n1_phot}}
\tablewidth{0pt}
\tablehead{
 & & & & \colhead{Fr. Limit} & & & \\
\colhead{Nova} & 
\colhead{MJD} & \colhead{$m_V$\tablenotemark{a}} & \colhead{Err($m_V$)} & \colhead{$m_V$} &
\colhead{MJD} & \colhead{$m_B$} & \colhead{Err($m_B$)} }

\startdata

NGC 205 1 &52986.92 & \nodata & \nodata &21.45 & \nodata & \nodata & \nodata \\
 &52987.63 & \nodata & \nodata &20.38 & \nodata & \nodata & \nodata \\
 &52988.63 & \nodata & \nodata &20.97 & \nodata & \nodata & \nodata \\
 &52989.64 & \nodata & \nodata &19.87 & \nodata & \nodata & \nodata \\
 &52990.64 & \nodata & \nodata &20.44 & \nodata & \nodata & \nodata \\
 &52991.63 & \nodata & \nodata &20.63 & \nodata & \nodata & \nodata \\
 &52992.63 &21.09 & 0.39 &20.68 & \nodata & \nodata & \nodata \\
 &52993.65 &21.01 & 0.35 &20.63 & \nodata & \nodata & \nodata \\
 &52994.66 &21.52 & 0.41 &21.08 & \nodata & \nodata & \nodata \\
 &52996.66 & \nodata & \nodata &19.04 & \nodata & \nodata & \nodata \\
 &52998.64 &20.71 & 0.20 &21.07 & \nodata & \nodata & \nodata \\
 &53000.67 & \nodata & \nodata &20.63 & \nodata & \nodata & \nodata \\
 &53001.66 & \nodata & \nodata &20.59 & \nodata & \nodata & \nodata \\
 &53002.64 &20.45 & 0.22 &20.50 & \nodata & \nodata & \nodata \\
 &53009.63 &20.23 & 0.25 &20.17 & \nodata & \nodata & \nodata \\
 &53010.63 &20.23 & 0.31 &19.99 & \nodata & \nodata & \nodata \\
 &53011.63 &19.99 & 0.13 &20.54 & \nodata & \nodata & \nodata \\
 &53013.64 &20.19 & 0.17 &20.55 & \nodata & \nodata & \nodata \\
 &53016.64 &20.24 & 0.17 &20.73 & \nodata & \nodata & \nodata \\
 &53017.64 &20.02 & 0.11 &20.93 & \nodata & \nodata & \nodata \\
 &53021.65 &20.01 & 0.13 &20.63 & \nodata & \nodata & \nodata \\
 &53022.64 &20.04 & 0.13 &20.84 & \nodata & \nodata & \nodata \\
 &53023.64 & \nodata & \nodata &19.59 & \nodata & \nodata & \nodata \\
 &53024.64 & \nodata & \nodata &17.17 & \nodata & \nodata & \nodata \\
 &53031.66 & \nodata & \nodata &19.46 & \nodata & \nodata & \nodata \\
 &53032.63 &19.47 & 0.21 &19.68 & \nodata & \nodata & \nodata \\
 &53033.64 & \nodata & \nodata &16.59 & \nodata & \nodata & \nodata \\
 &53034.61 & \nodata & \nodata &19.33 & \nodata & \nodata & \nodata \\
 &53035.61 & \nodata & \nodata &19.57 & \nodata & \nodata & \nodata \\
 &53037.62 & \nodata & \nodata &19.05 & \nodata & \nodata & \nodata \\
 &53042.61 &20.28 & 0.26 &20.11 & \nodata & \nodata & \nodata \\
 &53043.61 &20.18 & 0.18 &20.63 & \nodata & \nodata & \nodata \\
 &53044.61 &20.22 & 0.17 &20.78 & \nodata & \nodata & \nodata \\
 &53045.61 &20.61 & 0.20 &21.02 & \nodata & \nodata & \nodata \\
 &53046.62 &20.39 & 0.20 &20.83 & \nodata & \nodata & \nodata \\
 & \nodata  & \nodata & \nodata & \nodata & 53047.61 & 21.50 & 0.24 \\
 & \nodata  & \nodata & \nodata & \nodata & 53048.60 & $>$21.10 & 0.21 \\
 &53049.60 &20.54 & 0.23 &20.97 & \nodata & \nodata & \nodata \\
 &53050.60 & \nodata & \nodata &19.26 & \nodata & \nodata & \nodata \\
 &53051.60 &20.39 & 0.18 &21.06 & \nodata & \nodata & \nodata \\
 &53052.60 &20.50 & 0.23 &20.68 & \nodata & \nodata & \nodata \\
 &53053.60 &20.52 & 0.20 &20.96 & \nodata & \nodata & \nodata
\enddata
\end{deluxetable}

\clearpage

\begin{deluxetable}{lrcccccccc}
\rotate
\tabletypesize{\small}
\tablecaption{Light Curve Properties\label{tab_nov_prop}}
\tablewidth{0pt}
\tablehead{
& & & & & & & \multicolumn{3}{c}{Decline}\\
\cline{8-10} \\
& \colhead{Max} & \colhead{Min\tablenotemark{a}} & \colhead{Ampl.} & & \colhead{Baseline} & \colhead{Rise Time} & 
\colhead{Baseline} & \colhead{Pts} & \colhead{Rate}\\
\colhead{Object} & \colhead{($m_V$)} & \colhead{($m_V$)} &
\colhead{($m_V$)} & \colhead{$<$B$-$V$>$} & \colhead{(days)} & 
\colhead{(days)} & \colhead{(days)} & \colhead{(N)} & 
\colhead{($m_V$ day$^{-1}$)} }

\startdata
M32 Nova 1	&$<$ 17.3 &	$>$ 26.0 & $>$ 8.7 & 0.14$\pm$0.08 &37	&$<$ 6  &37	&22	&0.065\\
NGC 205 Nova Cand 1&19.47	  &	$>$ 24.1 & $>$ 4.6 & 1.04$\pm$0.15 &61	&$>$ 40	&21	&10	&0.046\\
\enddata
\tablenotetext{a}{Derived from archival {\it HST} observations, see
\S\ref{verify}}
\end{deluxetable}

\clearpage
\begin{deluxetable}{lccclcllc}
\tabletypesize{\small}
\rotate
\tablecolumns{9}
\tablecaption{Nova Rates\label{tab_nova_rate}}
\tablewidth{0pt}
\tablehead{
 & \colhead{Bulk Nova Rate\tablenotemark{a}} &
\colhead{V$_{T,0}$\tablenotemark{b}} & 
\colhead{(V - K)$_0$\tablenotemark{c}} & \colhead{K$_{V,(V-K)}$} &
\colhead{K$_{2MASS,0}$\tablenotemark{d}} & \colhead{(m$-$M)$_0$\tablenotemark{e}}&
\colhead{L$_K$} & \colhead{LSNR} \\
\colhead{Galaxy} & \colhead{(yr$^{-1}$)} & \colhead{(mag)} & \colhead{(mag)} &
\colhead{(mag)} & \colhead{(mag)} & \colhead{(mag)} &
\colhead{($10^{10} L_{\odot,K}$)} 
& \colhead{(yr$^{-1} [10^{10} L_{\odot,K}]^{-1}$)}
}

\startdata
M81	&33$^{+13}_{-8}$   & 6.57 & 3.17$\pm$0.1 & 3.40$\pm$0.1 &
3.802$\pm$0.018 & 27.80$\pm$0.08 & 12.4$\pm$1.5 & 2.6$^{+1.0}_{-0.6}$\\[7pt]
\tableline
\\
M32	&2$^{+2.4}_{-1.0}$ & 7.84 & 3.13$\pm$0.1 & 4.71$\pm$0.1 & 
5.072$\pm$0.017 & 24.43$\pm$0.1 & 0.167$\pm$0.023 & 12.0$^{+14.4}_{-6.0}$\\[7pt]
NGC 205	&2$^{+2.2}_{-1.0}$ & 7.97 & 2.12$\pm$0.2 & 5.85$\pm$0.2 &
5.564$\pm$0.045 & 24.60$\pm$0.3 & 0.0682$\pm$0.0267 & 29.3$^{+32.3}_{-14.7}$\\[7pt]
 &0$^{+1.5}_{-0.0}$\tablenotemark{f}
&\nodata&\nodata&\nodata&\nodata&\nodata&\nodata& $<$ 22\tablenotemark{f}\\[7pt]
NGC 147	&0$^{+2.0}_{-0.0}$ & 8.93 &\nodata&6.9$\pm$0.2\tablenotemark{g}&
7.137$\pm$0.063 & 24.39$\pm$0.05 & 0.0224$\pm$0.0047 & $<$ 89\\[7pt]
NGC 185	&0$^{+1.8}_{-0.0}$ & 8.55 &\nodata&6.3$\pm$0.2\tablenotemark{g}&
6.495$\pm$0.051 & 23.96$\pm$0.21 & 0.0269$\pm$0.0082 & $<$ 67\\[7pt]
\tableline
\\
LGdEs	&4$^{+4.2}_{-1.4}$ &\nodata&\nodata&\nodata&\nodata&
\nodata & 0.284$\pm$0.036  & 14.1$^{+14.8}_{-4.9}$\\[7pt]
     	&2$^{+3.9}_{-1.4}$\tablenotemark{f} &\nodata&\nodata&\nodata&\nodata&
\nodata &\nodata  & 7.0$^{+13.7}_{-4.9}$\tablenotemark{f}\\
\enddata
\tablenotetext{a}{Nova rate references: M81 - \citet{nei04},
others - this study}
\tablenotetext{b}{from \citet{dev91}}
\tablenotetext{c}{Color references: M81 - \citet{aar78}, others - \citet{fro78}}
\tablenotetext{d}{from \citet{jar03}, but corrected for reddening using the
formula A(K) $\simeq$ 0.085A(B) from \citet{sch98} with A(B) values from
\citet{dev91}}
\tablenotetext{e}{Distance references: M81 - \citet{fre01},
M32 - \citet{gri96}, NGC 205 - \citet{lee96}, NGC 147 - \citet{han97}, 
and NGC 185 - \citet{lee93}}
\tablenotetext{f}{These are the values assuming that the variable in NGC
205 is not a nova}
\tablenotetext{g}{Estimated from K$_{2MASS,0}$ and subtracting the 0.2
systematic offset found by \citet{fer03}}
\end{deluxetable}



\begin{thebibliography}{}
\bibitem[Aaronson(1978)]{aar78} Aaronson, M.\ 1978, \apjl, 221, L103
\bibitem[Arp(1956)]{arp56} Arp, H.~C.\ 1956, \aj, 61, 15
\bibitem[della Valle et al.(1994)]{del94} della Valle, M., Rosino, L., 
	Bianchini, A., \& Livio, M.\ 1994, \aap, 287, 403 
\bibitem[de Vaucouleurs et al.(1991)]{dev91} de Vaucouleurs, G.,
	de Vaucouleurs, A., Corwin, H.~G., Buta, R.~J., Paturel, G., \& 
	Fouque, P.\ 1991, Volume 1-3, XII, 2069 pp.~Springer-Verlag Berlin 
	Heidelberg New York
\bibitem[Ferrarese, C{\^ o}t{\' e}, \& Jord{\' a}n(2003)]{fer03} Ferrarese, 
	L., C{\^ o}t{\' e}, P., \& Jord{\' a}n, A.\ 2003, \apj, 599, 1302
\bibitem[Freedman et al.(2001)]{fre01} Freedman, W.~L.~et al.\ 2001, \apj, 
	553, 47
\bibitem[Frogel, Persson, Matthews, \& Aaronson(1978)]{fro78} 
	Frogel, J.~A., Persson, S.~E., Matthews, K., \& Aaronson, M.\ 1978, 
	\apj, 220, 75
\bibitem[Grillmair et al.(1996)]{gri96} Grillmair, C.~J.~et al.\ 1996, \aj, 
	112, 1975
\bibitem[Han et al.(1997)]{han97} Han, M., Hoessel, J.~G., Gallagher, J.~S., 
	Holtsman, J., \& Stetson, P.~B.\ 1997, \aj, 113, 1001
\bibitem[Jarrett et al.(2003)]{jar03} Jarrett, T.~H., Chester, T., Cutri, R., 
	Schneider, S.~E., \& Huchra, J.~P.\ 2003, \aj, 125, 525
\bibitem[de Jong et al.(2004)]{dej04} de Jong, J.~T.~A., et al.\ 2004, \aap, 
	417, 461
\bibitem[Lee, Freedman, \& Madore(1993)]{lee93} Lee, M.~G., Freedman, W.~L., 
	\& Madore, B.~F.\ 1993, \aj, 106, 964
\bibitem[Lee(1996)]{lee96} Lee, M.~G.\ 1996, \aj, 112, 1438
\bibitem[Magnier et al.(1992)]{mag92} Magnier, E.~A., Lewin, W.~H.~G., 
	van Paradijs, J., Hasinger, G., Jain, A., Pietsch, W., \& Truemper, 
	J.\ 1992, \aaps, 96, 379
\bibitem[Neill \& Shara(2004)]{nei04} Neill, J.~D.~\& Shara, M. M. 2004
	\aj, 127, 816
\bibitem[Nowotny el al.(2003)]{now03} Nowotny, W., Kerschbaum, F., Olofsson, 
	H., \& Schwarz, H.~E.\ 2003, \aap, 403, 93
\bibitem[Rosino(1973)]{ros73} Rosino, L.\ 1973, \aaps, 9, 347
\bibitem[Sandage \& Tammann(1981)]{san81} Sandage, A.~\& Tammann, G.~A.\ 1981, 
	Carnegie Inst.~of Washington, Publ.~635
\bibitem[Schlegel, Finkbeiner, \& Davis(1998)]{sch98} Schlegel, D.~J., 
	Finkbeiner, D.~P., \& Davis, M.\ 1998, \apj, 500, 525
\bibitem[Shafter, Ciardullo, \& Pritchet(2000)]{sha00} Shafter, A. W., 
	Ciardullo, R., \& Pritchet, C. J. 2000, ApJ,530, 193
\bibitem[Shafter \& Irby(2001)]{sha01} Shafter, A.~W.~\& Irby, B.~K.\ 2001, 
	\apj, 563, 749
\bibitem[Stetson(1987)]{ste87} Stetson, P.~B.\ 1987, \pasp, 99, 191
\bibitem[Tody(1986)]{tod86} Tody, D. 1986, Proc. SPIE, 627, 733
\bibitem[Uglesich et al.(2004)]{ugl04} Uglesich, R.~R., Crotts, A.~P.~S., 
	Baltz, E.~A., de Jong, J., Boyle, R.~P., \& Corbally, C.~J.\ 2004,
	\apj, 612, 877 
\end{thebibliography}
\end{document}